\documentclass[]{amsart}

\usepackage{graphicx}
\usepackage{subfigure}
\usepackage{mathptmx}      % use Times fonts if available on your TeX system
\usepackage{url}

\newtheorem{proposition}{Proposition}
\newtheorem{problem}{Problem}

\pdfoutput=1

\begin{document}

%\begin{frontmatter}

\title{Slime mould computes planar shapes}

\author{Andrew Adamatzky}

\address{University of the West of England, Bristol BS16 1QY, UK \\
              \url{andrew.adamatzky@uwe.ac.uk}           %  \\
}


\begin{abstract}
Computing a polygon defining a set of planar points is a classical problem of modern 
computational geometry. In laboratory experiments we demonstrate that a concave 
hull, a connected $\alpha$-shape without holes, of a finite planar set is 
approximated by slime mould \emph{Physarum polycephalum}. 
We represent planar points with sources of long-distance attractants and short-distance 
repellents and inoculate a piece of plasmodium outside the data set. The plasmodium moves 
towards the data and envelops it by pronounced protoplasmic tubes. 

\vspace{0.5cm}

\noindent
\emph{Keywords:} Unconventional computing, Physarum polycephalum, Concave hull, Concave hull, Alpha shape
\end{abstract}

\maketitle

\section{Introduction}
\label{intro}

A slime mould \emph{Physarum polycephalum} has a rich life cycle~\cite{stephenson_2000}, which includes fruit bodies, spores, single-cell amoebas, and syncytium.  Plasmodium is a vegetative stage of \emph{Physarum polycephalum}, syncytium, a single cell with  many nuclei. The plasmodium consumes microscopic particles. During its foraging behaviour the plasmodium 
spans scattered sources of nutrients with a network of  protoplasmic tubes. The protoplasmic network is optimised to cover all sources of food and to deliver robust and speedy transportation of nutrients and metabolites in the plasmodium body. 
Plasmodium's foraging behaviour can be interpreted as computation, when data are represented by spatial configurations of attractants and repellents, and  results by structures of protoplasmic 
network~\cite{adamatzky_naturewissenschaften_2007,adamatzky_physarummachines}.
Plasmodium satisfactory solves many computational problems with natural parallelism, including
shortest path~\cite{nakagaki_2000,nakagaki_2001a}, 
implementation of storage modification machines~\cite{adamatzky_ppl_2007},
Voronoi diagram~\cite{shirakawa}, 
logical computing~\cite{tsuda_2004,adamatzky_gates}, 
process algebra~\cite{schumann_adamatzky_2009}; see overview in~\cite{adamatzky_physarummachines}.

A convex hull of a finite planar set is a smallest region of the plane containing the set~\cite{preparata_shamos}. Classical 
algorithms of convex hull construction include Jarvis' gift wrapping~\cite{jarvis_1973},
Preparata-Hong's divide and conquer~\cite{preparata_hong}, Graham's scan of pre-sorted set 
of points~\cite{graham_1972}, and Akl-Toussaint's quick algorithm based on removing points lying 
inside a convex quadrilateral of extreme points~\cite{akl_toussaint}; see reviews and discussion of 
the algorithms in~\cite{preparata_shamos,deberg_2008} and codes of implementations in~\cite{orourke}.
Multi-processor and systolic-processor solutions of the convex hull problem are mostly 
based on parallelisation of the classical serial algorithms~\cite{day_tracey_1998,evans_mai_1985} or
exploring extremal properties of the given set~\cite{hayashi:nakano,kim:stojmenovic,dehne:hassenklover};
see also review in~\cite{chen_nakano_wada_2000}. Cellular-automata algorithm are based on propagating 
patterns halting their growth in the hull's boundaries~\cite{adamatzky_1994} or encapsulating the 
data set and developing synchronisation patterns~\cite{torbey_akl_2008,clarridge_salomaa_2008}, are theoretical 
precursors for bio-inspired approach to computation of convex hulls.

A convex hull is economical but rough reconstruction of an object shape form the object's sample points. 
It often lacks details necessary for realistic visualisation of objects. In 1983 Edelsbrunner, Kirkpatrick and Seidel~\cite{edelsbrunner_1983} introduced $\alpha$-shapes, generalisation of convex hulls, where control parameter $\alpha$ allows for a smooth transition between crude and fine approximations of point sets. There are hundreds of algorithms for computing $\alpha$-shapes yet no experimental laboratory prototypes of chemical, physical or biological computers for approximation of $\alpha$-shapes. Present paper partly fills the gap and shows how to approximate $\alpha$-shapes without holes using foraging behaviour of \emph{P. polycephalum}.

\section{Experimental}
\label{experimental}

\begin{figure}[!tbp]
\centering
\subfigure[]{\includegraphics[width=0.42\textwidth]{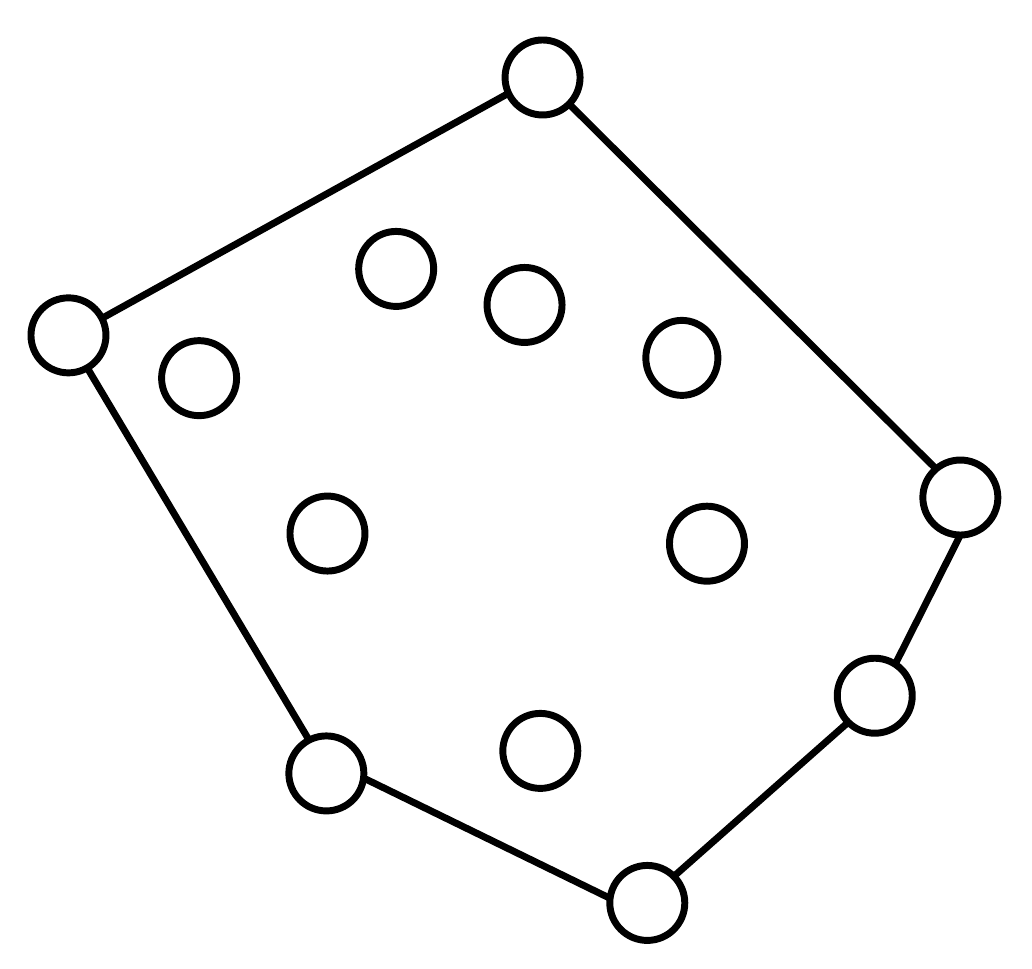}} %a
\subfigure[]{\includegraphics[width=0.42\textwidth]{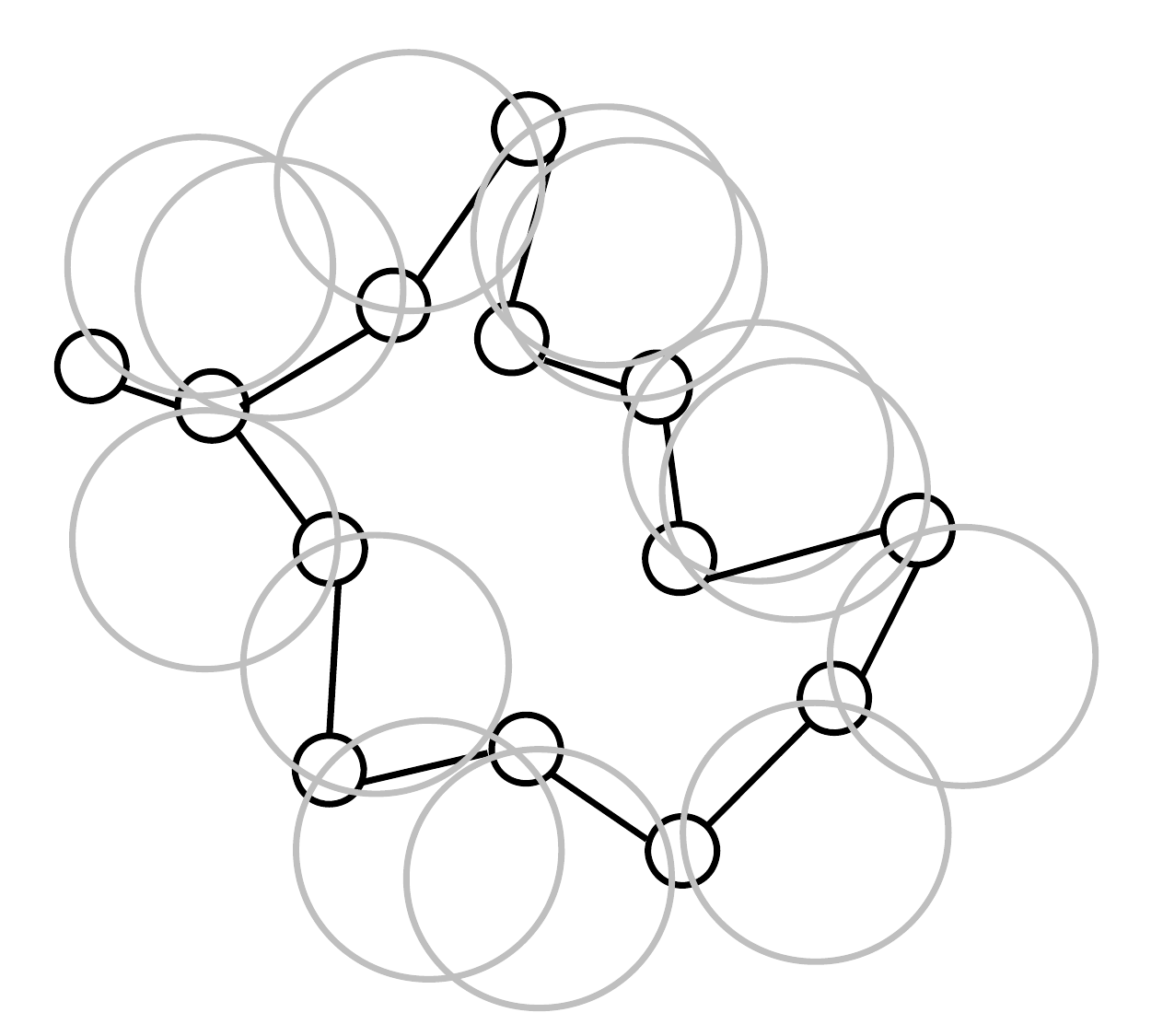}} %b
\subfigure[]{\includegraphics[width=0.42\textwidth]{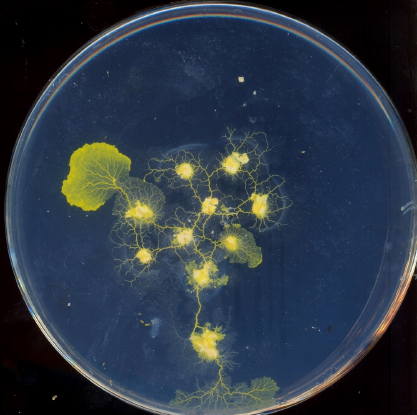}}         %c 
\subfigure[]{\includegraphics[width=0.42\textwidth]{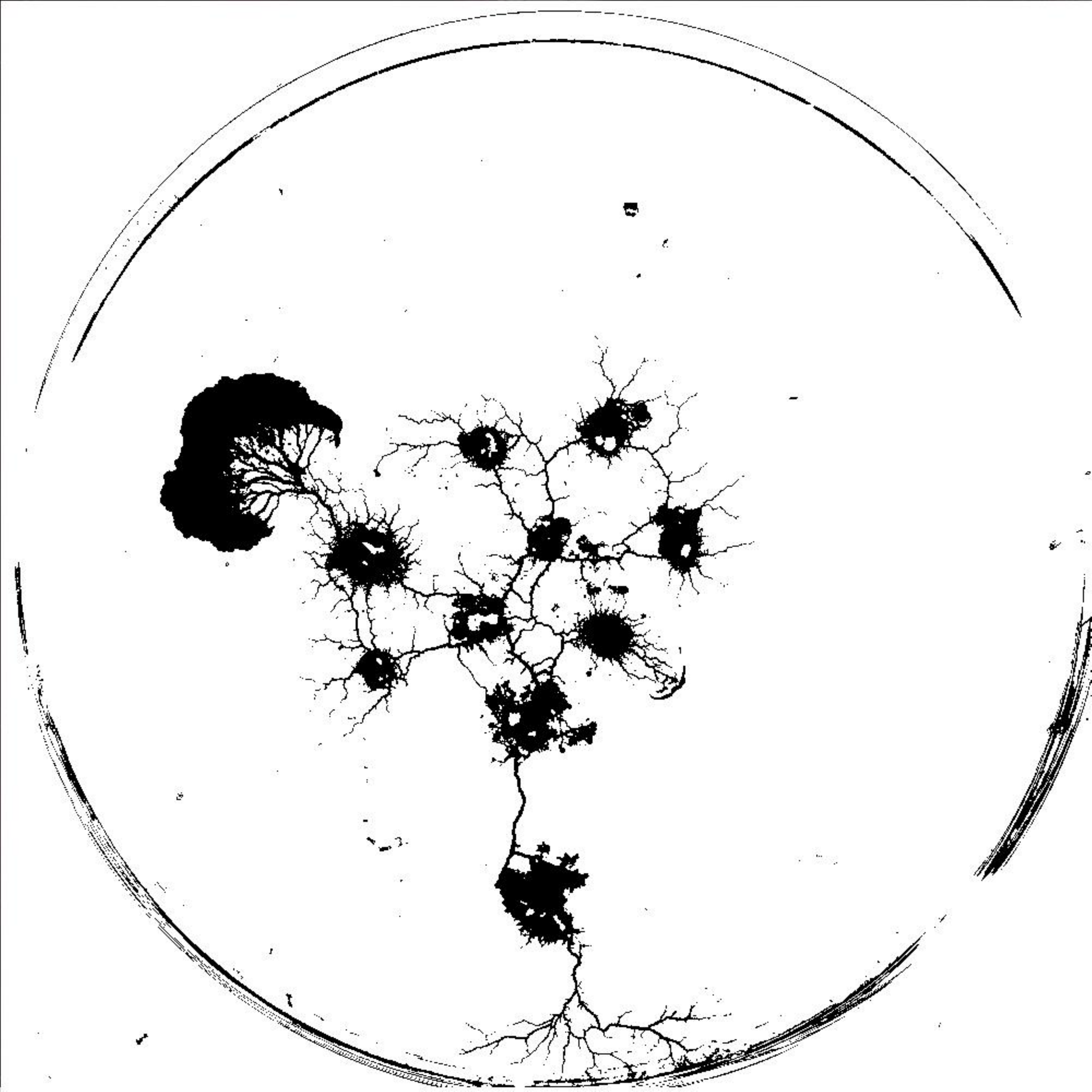}}   %d
\subfigure[]{\includegraphics[width=0.42\textwidth]{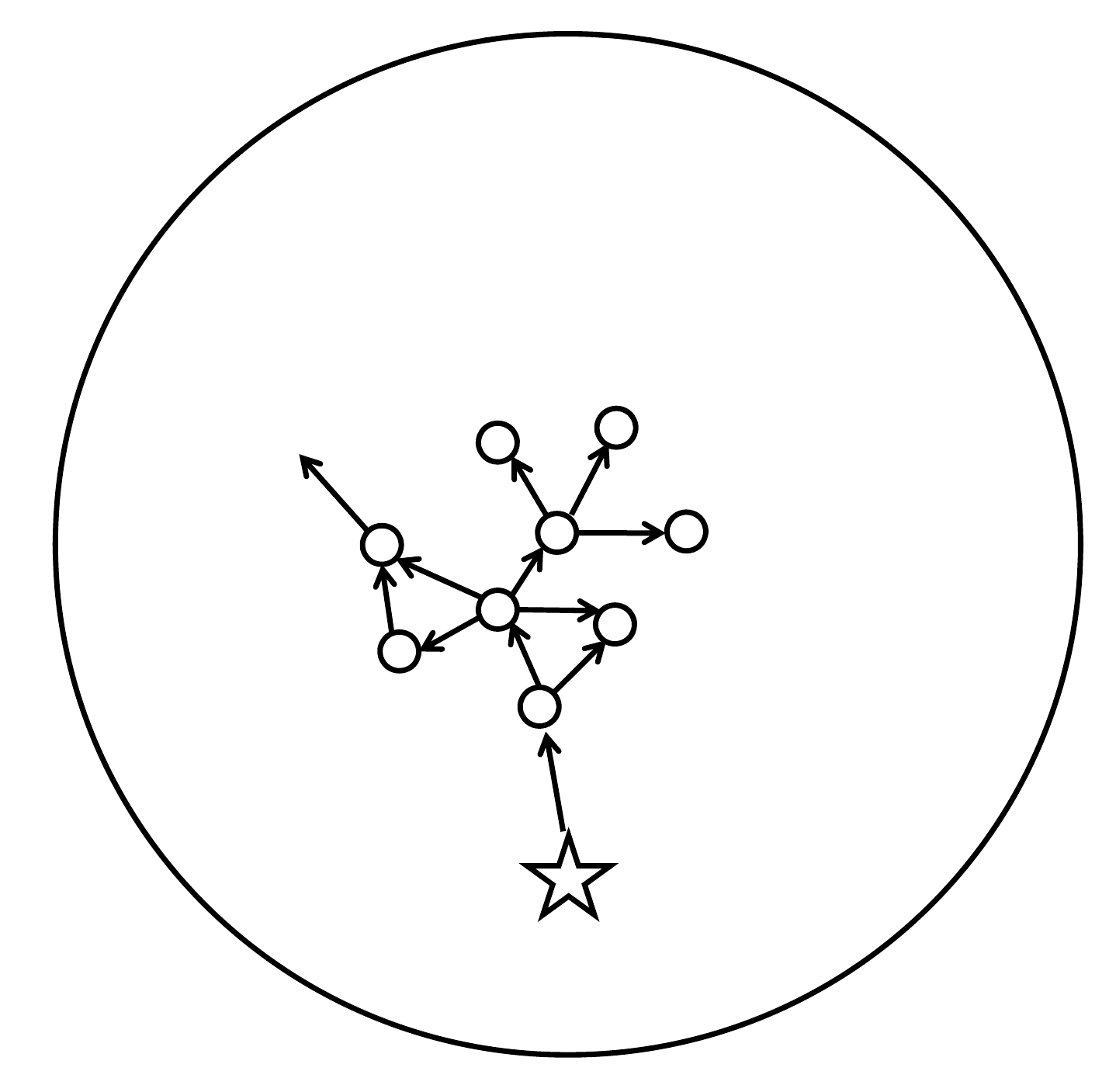}} %e
\caption{Basics. (a)~Example of convex hull. (b)~Example of concave hull. Points of set $\mathbf P$ are empty discs.
 (c--d)~Spanning oat flakes, representing $\mathbf P$, by a network of protoplasmic tubes:
 scanned image (c) and its binarisation (d). 
 (e)~Scheme of the plasmodium propagation, circles are points, or oat flakes, of $\mathbf P$ and 
 star marks site of inoculation, arrows are protoplasmic tubes.
}
\label{basics:abcde}
\end{figure}

A convex hull  of a finite set  $\mathbf P$  of planar points is the smallest convex 
polygon that contains all points  of $\mathbf P$ (Fig.~\ref{basics:abcde}a). $\alpha$-hull of $\mathbf P$ 
is an intersection of the complement of all closed discs of radius $1/\alpha$ that includes no points 
of $\mathbf P$~\cite{edelsbrunner_1983,edelsbunner_1994}. $\alpha$-shape is a convex hull  
when $\alpha \rightarrow \infty$. With decrease of $\alpha$ the shapes may shrink, develop holes 
and become disconnected, the shapes collapse to $\mathbf P$  when $\alpha \rightarrow 0$. 
A concave hull is  non-convex polygon representing area occupied by $\mathbf P$. A concave hull is a connected
$\alpha$-shape without holes (Fig.~\ref{basics:abcde}b).

\begin{problem}
Given planar set $\mathbf P$ represented by physical objects plasmodium of \emph{P. polycephalum} must represent
concave hull of $\mathbf P$ by its largest protoplasmic tube.
\end{problem}

We cultivate \emph{P. polycephalum} in plastic containers, on paper kitchen towels sprinkled with
still drinking water and fed with oat flakes. For experiments we use polystyrene Petri dishes (round, diameter
120~mm, and rectangular $120 \times 120$~mm) and  2\% agar gel (Select agar, Sigma Aldrich) as a 
non-nutrient substrate.  Images of plasmodium are recorded by scanning Petri dishes in Epson Perfection 4490. Photos are taken using FujiPix 6000 camera.

\section{Results}
\label{results}

The fist algorithm of convex hull construction~\cite{jarvis_1973} was based on cognitive tactic techniques 
we use in our everyday's life. We select a starting point which is extremal point of $\mathbf P$. We pull 
a rope (anti-)clockwise to other extremal point. We continue until the set $\mathbf P$ is wrapped 
completely. The computation stops when we reach the starting point. Let we represent data points $\mathbf P$ by sources of attractants only, e.g. by oat flakes (Fig.~\ref{basics:abcde}cde). We place a piece of plasmodium at some distance away from the set of points, see scheme in Fig.~\ref{basics:abcde}e. The plasmodium propagates towards 
set $\mathbf P$, colonises oat flakes (Fig.~\ref{basics:abcde}cd) and spans them with a network of protoplasmic tubes 
(Fig.~\ref{basics:abcde}e). No hull is constructed. When all data points $\mathbf P$ are colonised and spanned by protoplasmic network the plasmodium ventures to explore the space around $\mathbf P$.  

\begin{proposition}
P. polycephalum does not compute concave or convex hull of a set represented by attracting sources. 
\end{proposition}

\begin{figure}[!tbp]
\centering
\subfigure[]{\includegraphics[width=0.42\textwidth]{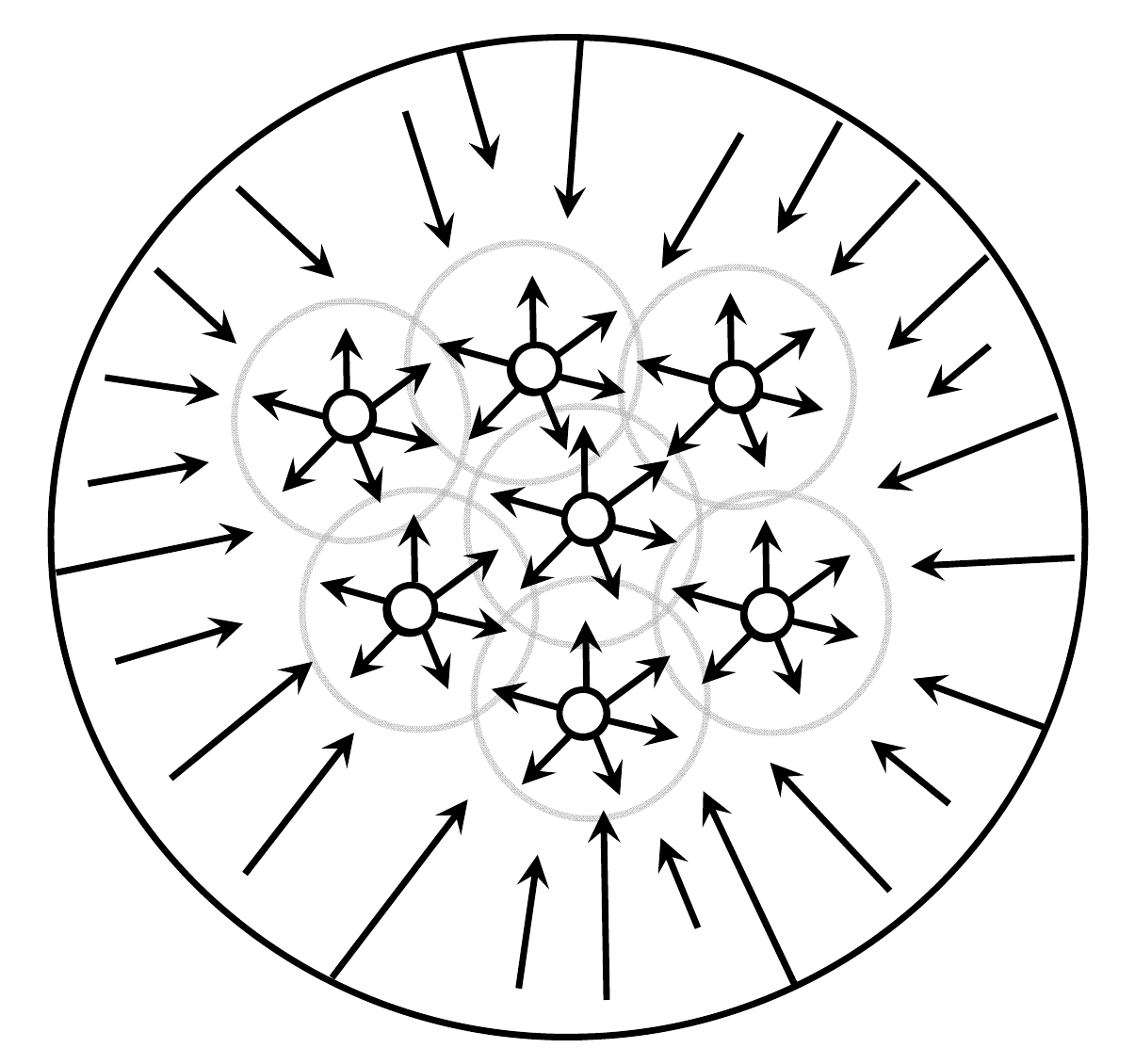}} \\ %f --> a 
\subfigure[]{\includegraphics[width=0.42\textwidth]{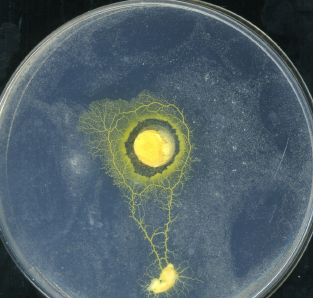}} %g --> b
\subfigure[]{\includegraphics[width=0.42\textwidth]{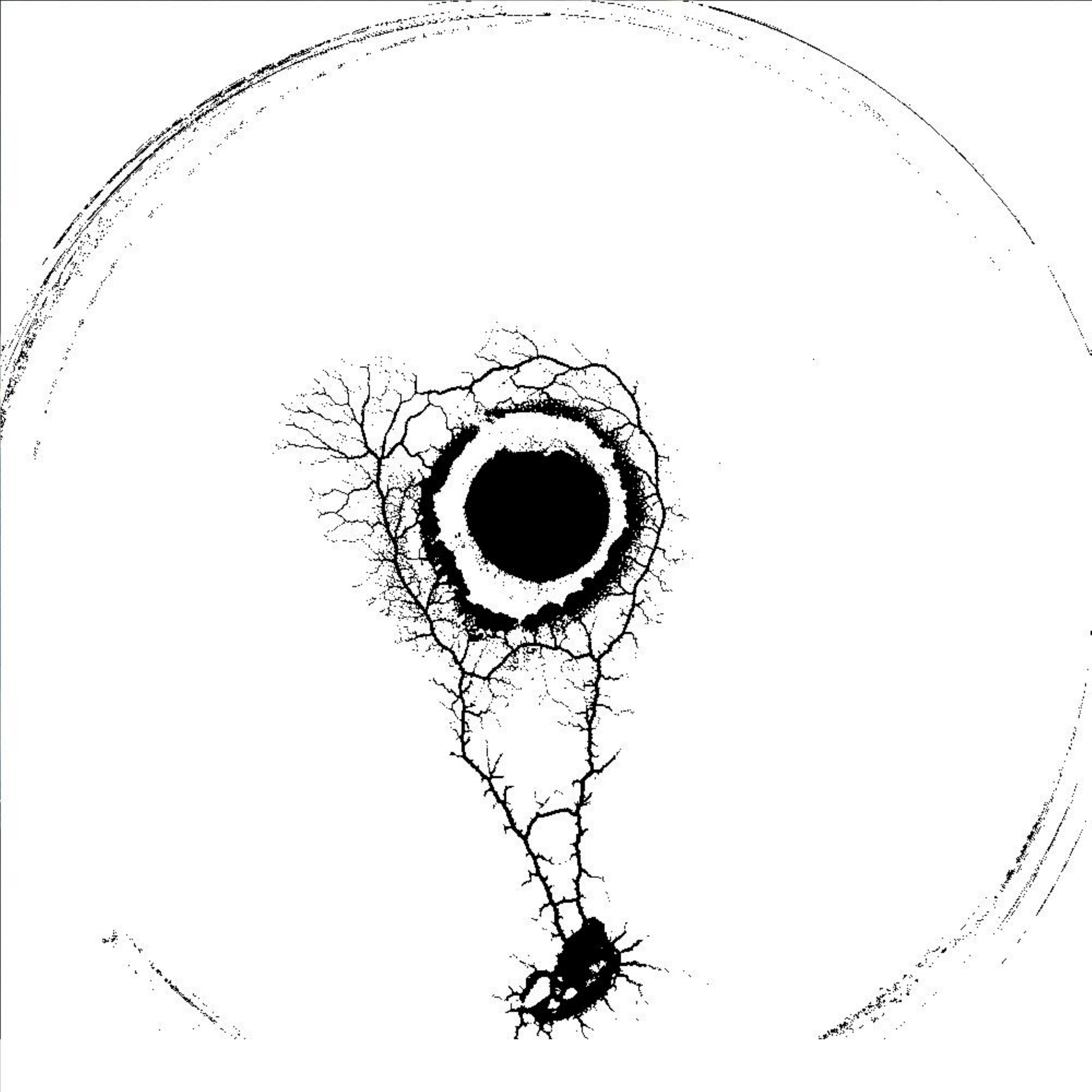}} %h --> c
\subfigure[]{\includegraphics[width=0.6\textwidth]{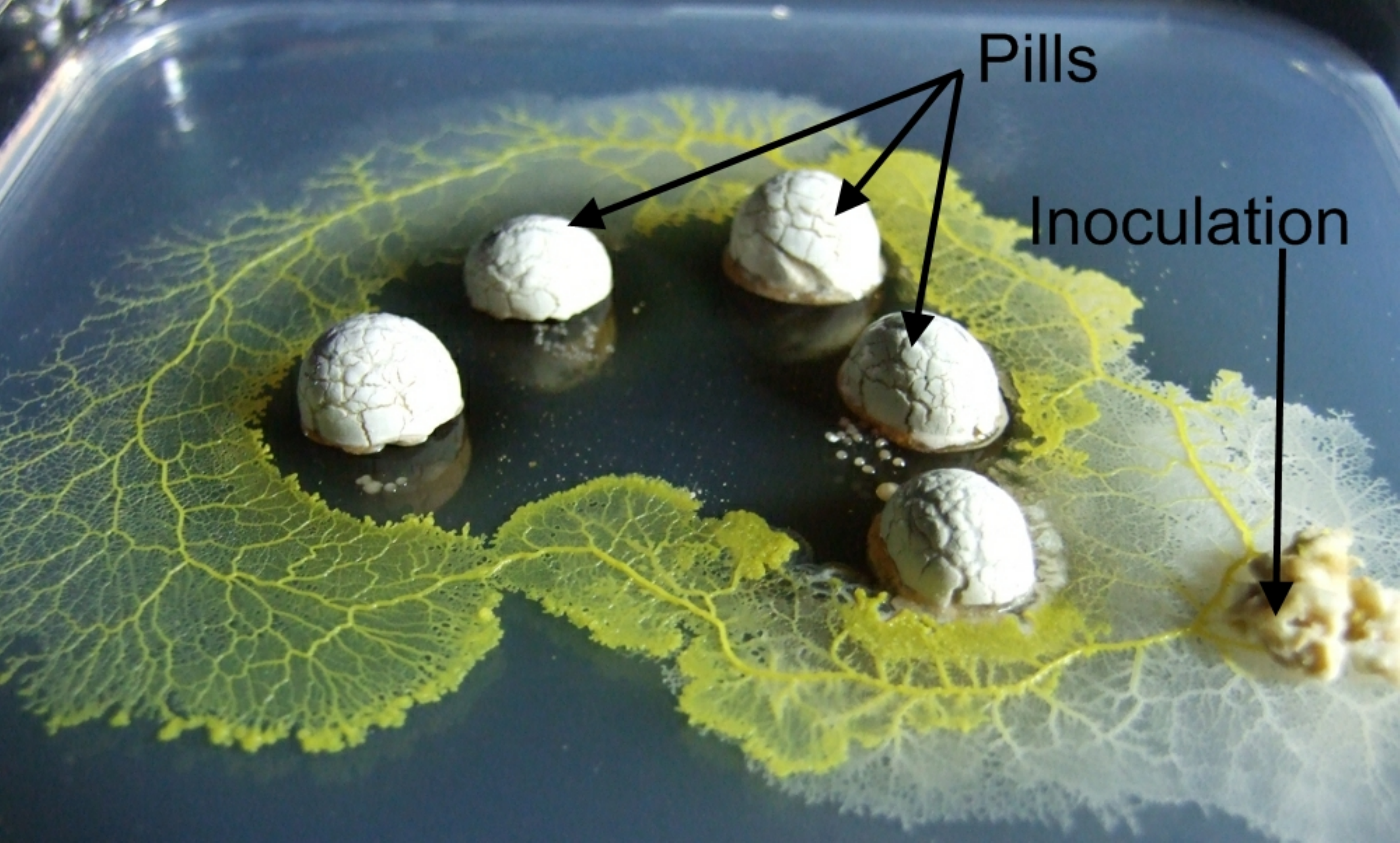}} %i -->d 
\caption{Experimental. (a)~Proposed distribution of attracting and repelling gradients which may 
force plasmodium to approximate a concave hull (arrows aiming towards set of discs are attractive forces, arrows originating in data points, discs, are repelling forces). 
(bc)~Plasmodium's interaction with the single half-pill of Kalms Tablets: scanned (b) and binarised (c) images;
at the beginning of experiment plasmodium is inoculated in southmost part of Petri dish (pill of Kalms Tablets is at the centre of the Petri dish).
(d)~Typical experimental setup, pills representing $\mathbf P$ and the initial site of plasmodium's inoculation are shown by arrows.}
\label{basics:fghi}
\end{figure}

Using repellents only, e.g. as a set of obstacles between attractants and inoculation site, will do no good because plasmodium will just pass around the repellents and leave (see details in \cite{adamatzky_physarummachines}, controlling Physarum with salt). The only solution would be to employ attractants to `pull' plasmodium towards planar set $\mathbf P$ and to use repellents to prevent plasmodium from spanning the points of $\mathbf P$ (Fig.~\ref{basics:fghi}a). Strength of repellents should be proportional to $\alpha$ and thus will determine exact shape of the constructed hull. This corresponds to original definition~\cite{edelsbrunner_1983} that
$\alpha$-hull of $\mathbf P$ is the intersection of all closed discs with radius $l/\alpha$ that contain all the points
of $\mathbf P$.

\begin{proposition}
Plasmodium of P. polycephalum approximates connected $\alpha$-hull (without holes) 
of a finite planar set, which points are represented by sources of long-distance 
attractants and short-distance repellents. 
\end{proposition}

To prove the proposition constructively we must find an appropriate representation of $\mathbf P$ and demonstrate
in experiments viability of the approach. While experimenting with different potential candidates for combined representation of attractants and repellents we found that the plasmodium's reaction to pills of Kalms Tablets and Kalms Sleep\footnote{G.~R.~Lane Health Products Limitted, Gloucester GL2 0GR, UK} is somewhat unusual. When presented with a half-pill of the Kalms Tablets/Sleep the plasmodium propagates towards the pill and forms, with its protoplasmic tubes, a circular enclosure around the pill (Fig.~\ref{basics:fghi}bc). Such a unique behaviour of plasmodium in presence of Kalms Tablets/Sleep indicates that a plasmodium could implement Jarvis's Gift Wrapping algorithm~\cite{jarvis_1973}, adapted to concave hulls without holes, if points of $\mathbf P$ are represented by the pills.

\begin{figure}[!tbp]
\centering
\subfigure[$t=12$~h]{\includegraphics[width=0.42\textwidth]{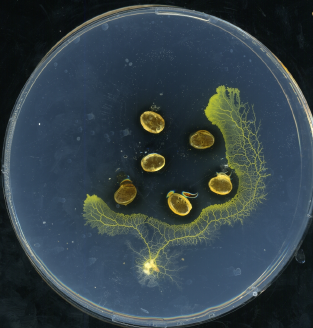}} %a 
\subfigure[$t=12$~h]{\includegraphics[width=0.42\textwidth]{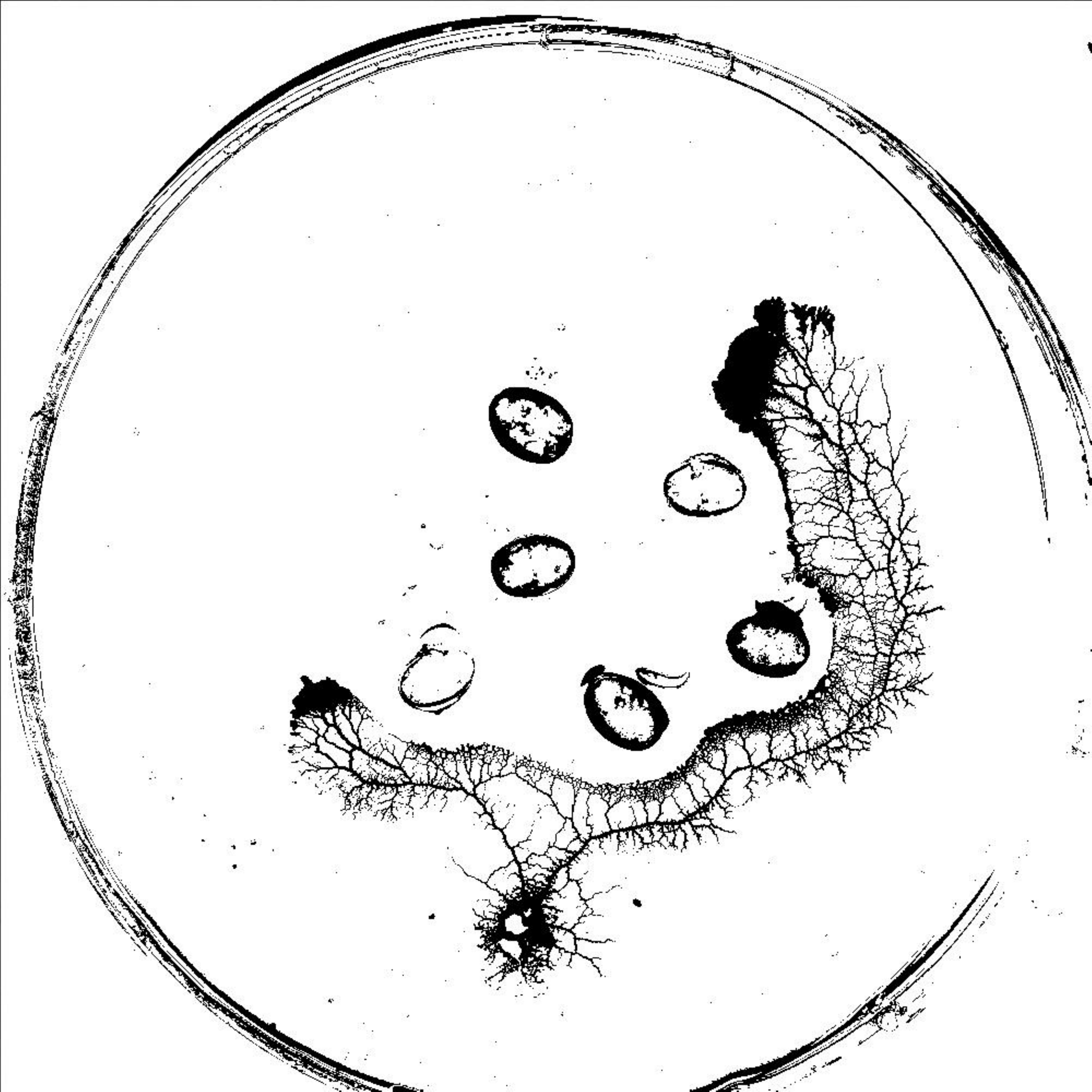}} %b
\subfigure[$t=24$~h]{\includegraphics[width=0.42\textwidth]{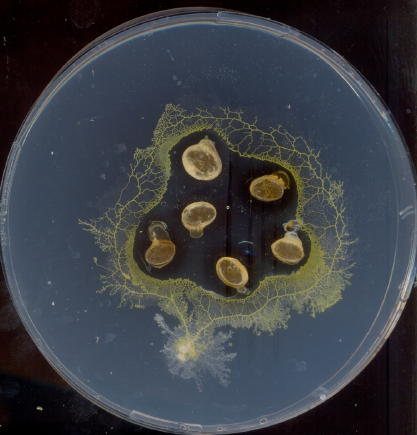}} %c
\subfigure[$t=24$~h]{\includegraphics[width=0.42\textwidth]{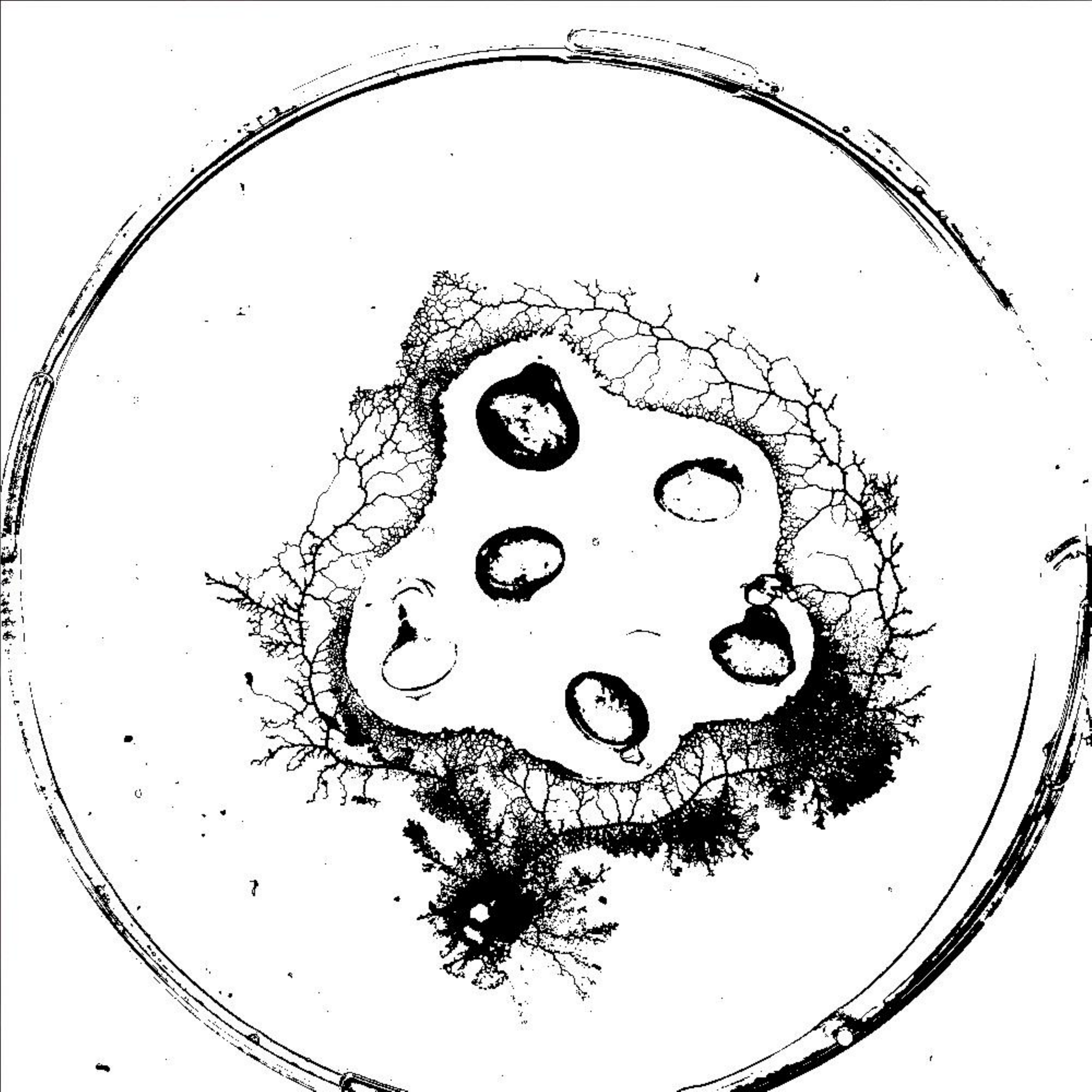}} %d
\subfigure[]{\includegraphics[width=0.42\textwidth]{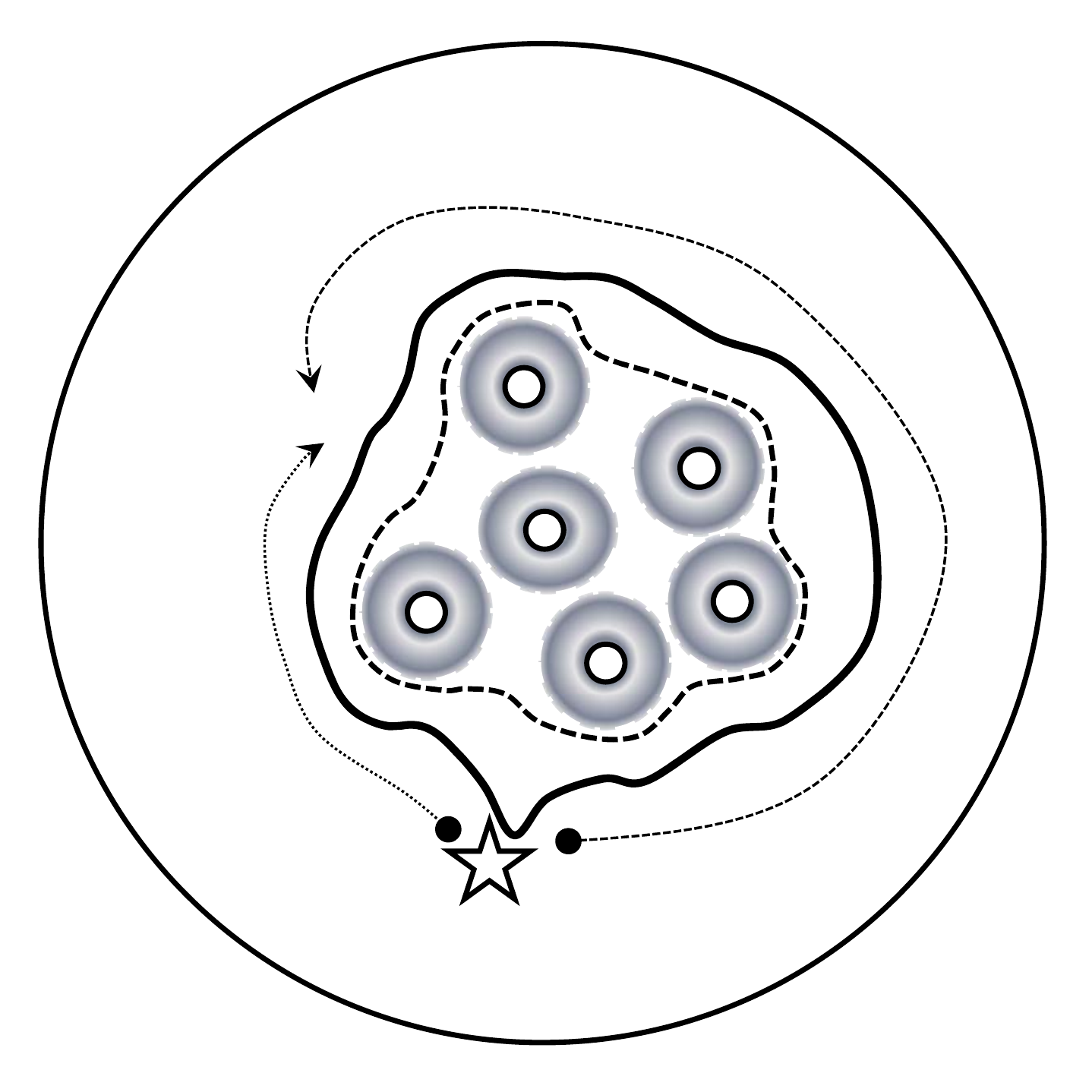}} %e
\caption{Computation of concave hulls of point set $\mathbf P$. Images~(ac) and 
their binarizations~(bd) taken 12~h~(ab) and 24~h~(cd) hours after inoculation. 
(e)~Scheme of the plasmodium interaction with $\mathbf P$: points/pills of $\mathbf P$ 
are circles; diffusing repellents, visible as black halo's surrounding pills in (a) and (c), 
are shown by grey gradients; feeding boundary of protoplasm is marked by dotted 
line; major protoplasmic tube, which approximates $CH({\mathbf P})$, is shown by solid line.}
\label{example:abcdef}
\end{figure}

We tested feasibility of the idea in 25 experiments. All experiments were successful. In each experiment we arranged 4-8 half-pills (they represented given planar set $\mathbf P$) in a random fashion near centre of a Petri dish and inoculated an oat flake colonised by plasmodium 2-4~cm away from the set $\mathbf P$. A typical experiment is illustrated in Fig.~\ref{example:abcdef}a--d. In 12~h after inoculation plasmodium propagates towards set $\mathbf P$ and starts enveloping the set with its body and network of protoplasmic tubes (Fig.~\ref{example:abcdef}ab). The plasmodium completes approximation of a shape by entirely enveloping $\mathbf P$ in next 12~h (Fig.~\ref{example:abcdef}cd). The plasmodium does not propagate inside configuration of pills (Fig.~\ref{example:abcdef}e).
  
\begin{figure}[!tbp]
\centering
\includegraphics[width=0.5\textwidth]{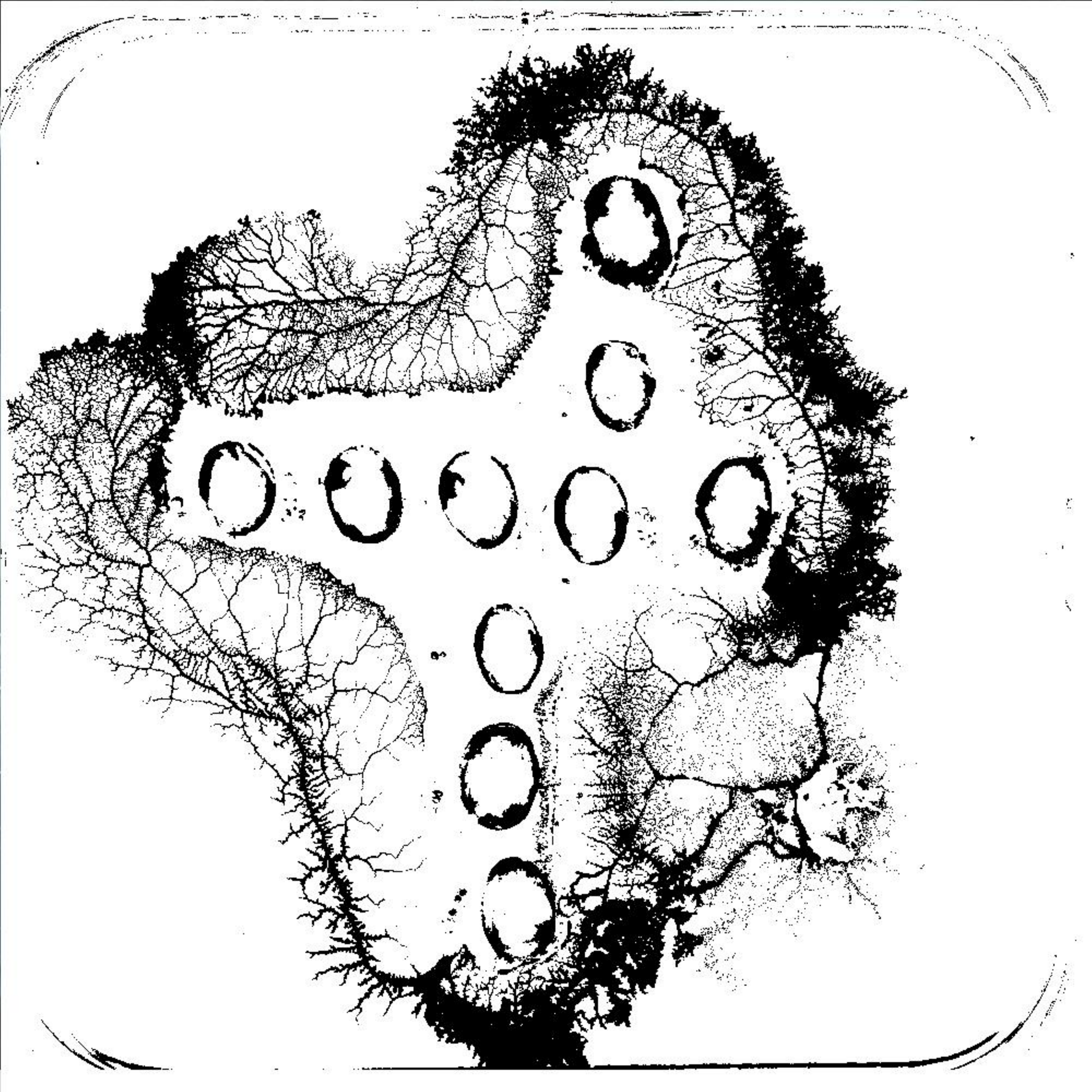}
\caption{Binarized image of plasmodium approximating concave hull of $\mathbf P$.}
\label{concave}
\end{figure}

\begin{figure}[!tbp]
\centering
\includegraphics[width=0.8\textwidth]{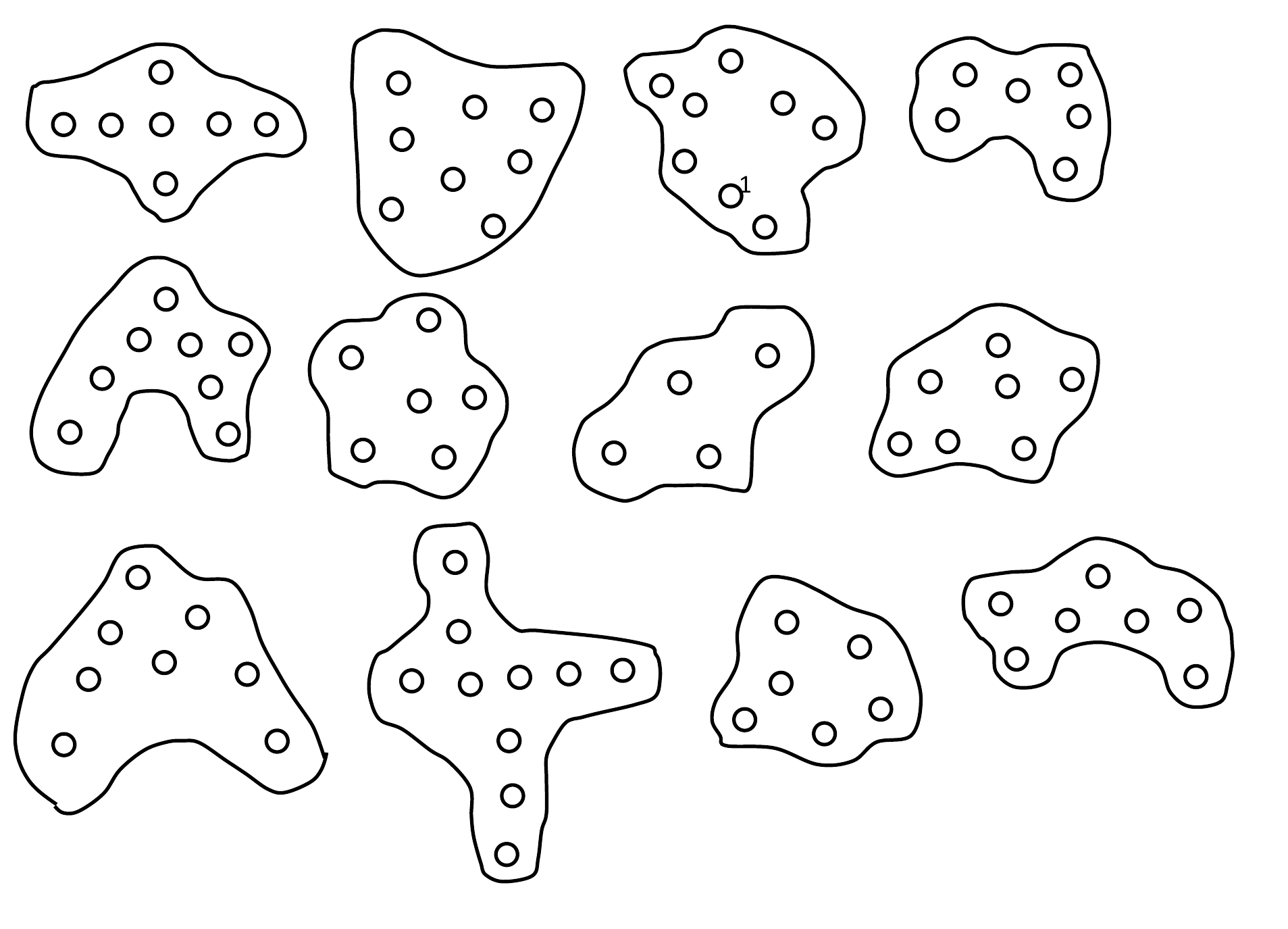}
\caption{Contours of convex and concave hulls constructed on small sets of planar points in laboratory experiments.}
\label{typicalhulls}
\end{figure}

Configuration of $\mathbf P$ in Fig.~\ref{example:abcdef}a--d favours approximation of a convex hull. If spatial configuration of points curves inwards then concave hull is approximated (Fig.~\ref{concave}). Typical hulls approximated by plasmodium of \emph{P. polycephalum} in our experiments are shown in Fig.~\ref{typicalhulls}. 
Hulls constructed by plasmodium match their counterparts calculated by classical algorithms.

Let us verify experimental laboratory results with computer simulation. A profile of plasmodium's 
propagating zone is isomorphic to shapes of wave-fragments in sub-excitable 
media~\cite{adamatzky_physarummachines}. When active zone of \emph{P. polycephalum} 
propagates two processes occur simultaneously --- propagation of the wave-shaped tip of the pseudopodium and 
formation of the trail of protoplasmic tubes. We simulate the (chemo-)tactic traveling of plasmodium using two-variable Oregonator equation~\cite{field_noyes_1974,tyson_fife}:
$$\frac{\partial u}{\partial t} = \frac{1}{\epsilon} (u - u^2 - (f v + \phi)\frac{u-q}{u+q}) + D_u \nabla^2 u$$
$$\frac{\partial v}{\partial t} = u - v .$$
The variable $u$ is abstracted as a local density of plasmodium's protoplasm and $v$ reflects local concentration of 
metabolites and nutrients. We integrate the system using Euler method with five-node Laplasian operator, time step $\Delta t=5\cdot10^{-3}$ and grid point spacing $\Delta x = 0.25$, with the following parameters: $\phi=\phi_0 - \eta/2$, $A=0.0011109$, $\phi_0=0.0766$, $\epsilon=0.03$, $f=1.4$, $q=0.022$.

 \begin{figure}[!tbp]
\centering
\subfigure[]{\includegraphics[width=0.48\textwidth]{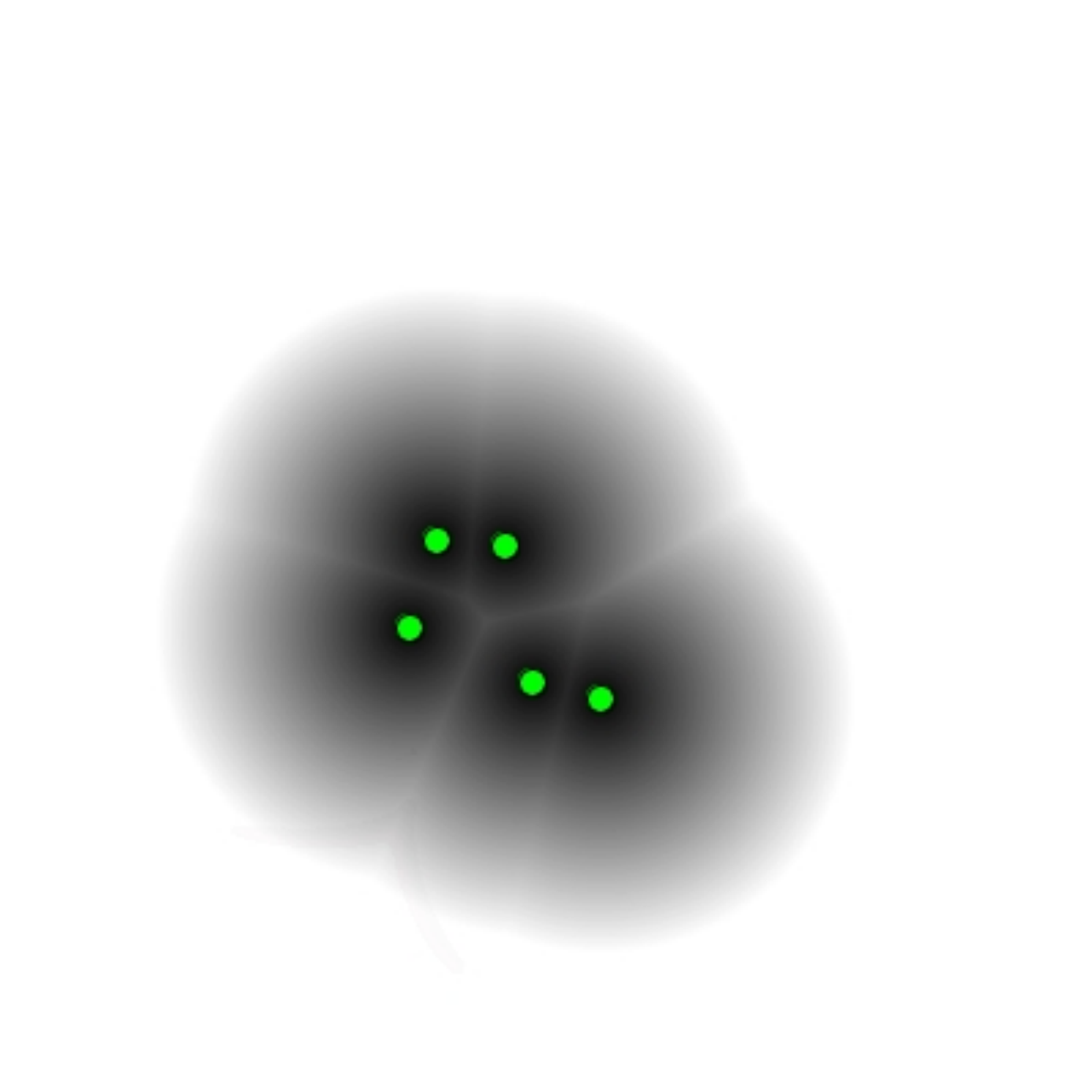}}
\subfigure[]{\includegraphics[width=0.48\textwidth]{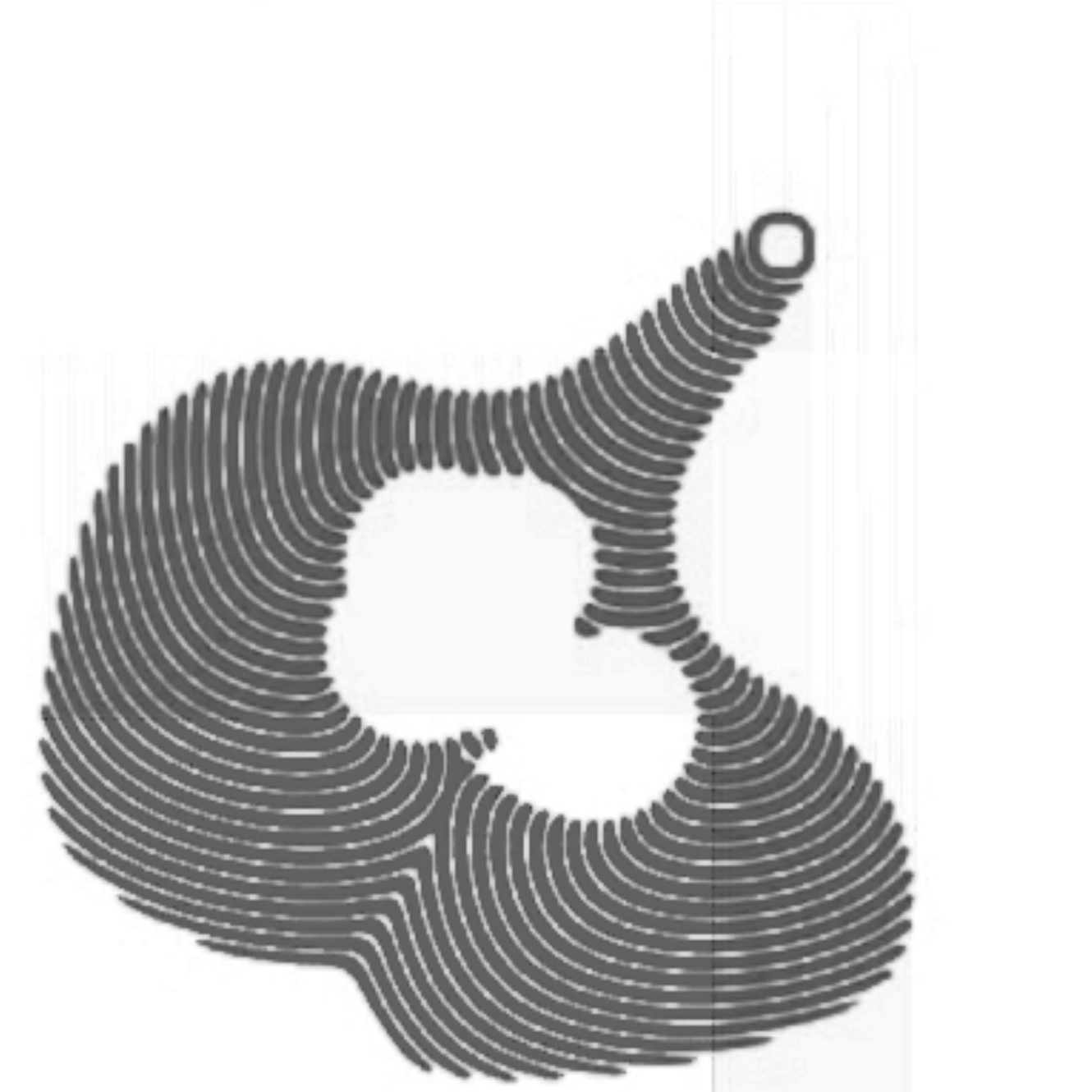}}
\subfigure[]{\includegraphics[width=0.48\textwidth]{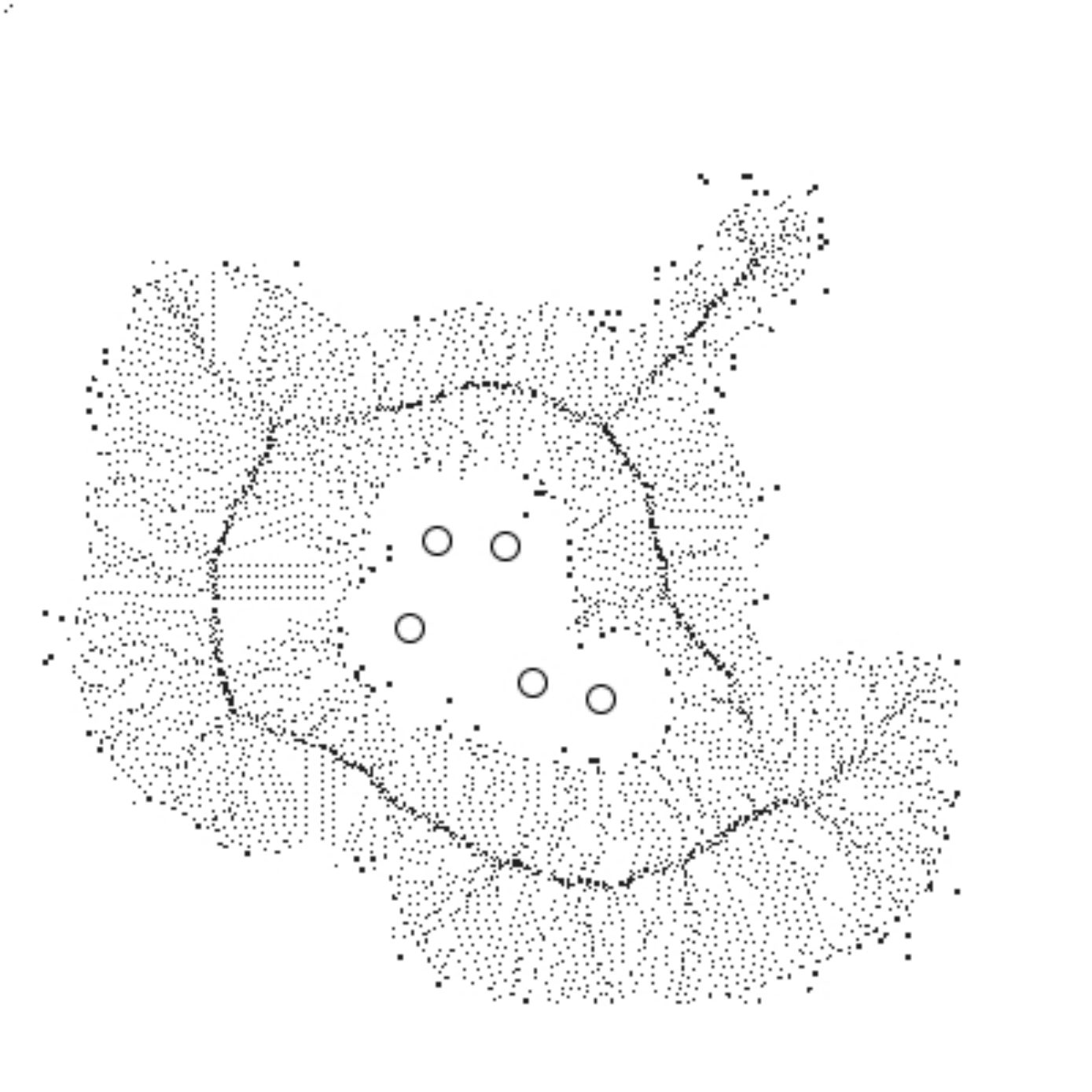}}
\caption{Simulating plasmodium approximation of convex hull of $\mathbf P$ in Oregonator model.
(a)~Gradients of repellents generated by points of $\mathbf P$. (b)~Time lapsed images of propagating plasmodium patterns. (c)~Structure of protoplasmic tubes developed.
}
\label{oregonator}
\end{figure}

Parameters $q$ and $f$ are inherited from model of Belousov-Zhabotinsky medium,
$\phi$ is proportional to local concentration of attractants and repellents. The parameter
 $\eta$ corresponds to a gradient of chemo-attractants emitted by data planar points. 
The medium is perturbed by an initial excitation, where a $11\times11$ sites are assigned $u=1.0$ each. The perturbation generates a  wave-fragment alike propagation of protoplasm which  travels along gradient $\eta$.  Repellents emitted by sites of $\mathbf P$ show very limited diffusion. Therefore repellents can be regarded as impassable obstacles (Fig.~\ref{oregonator}a).

To imitate formation of the protoplasmic tubes we store values of $u$ in matrix $\bf L$, which is processed at the end of simulation. For any site $x$ and time step $t$ if $u_x>0.1$ and $L_x=0$ then $L_x=1$. The matrix $\bf L$ represents time lapse superposition of 
propagating wave-fronts (Fig.~\ref{oregonator}b). The simulation is considered completed when propagating pattern envelops $\mathbf P$ and halts any further motion.  At the end of simulation we repeatedly apply the erosion operation~\cite{adamatzky_physarummachines} (which represents  a stretch-activation effect~\cite{kamiya_1959} necessary for formation of plasmodium tubes) to $\bf L$. The resultant protoplasmic network (Fig.~\ref{oregonator}c)  provides a good phenomenological match for networks recorded in laboratory experiments.

\section{Discussion}
\label{discussion}

In last five years we witnessed a substantial progress in design and fabrication of working prototypes of amorphous biological 
computing devices based on foraging behaviour of slime mould \emph{P. polycephalum}. Experimental laboratory prototypes of Physarum-based processors are developed for computation of minimum spanning tree, relative neighbourhood graph, 
Delaunay triangulation,  and Voronoi diagram, see overview in~\cite{adamatzky_physarummachines}. By present experimental demonstration of planar shape (convex and concave hulls) computing by  \emph{P. polycephalum} we draw a logic conclusion of the years of studies into computational properties of the slime mould with regards to problems of computational geometry. 

We used Kalms Tablet and Kalms Sleep pills (both brands cause similar effects to plasmodium behaviour) to represent data points. We exploited two properties of the pills. First, the pills contain Valerian and hops extracts, which are long distant attractants 
for \emph{P. polycephalum}~\cite{adamatzky:valerian} and thus `pull' plasmodium towards the date set. Second, the pills contain
short-distance, slow diffusing components (potential candidates are  magnesium stearate, stearid acid
and  titanium dioxide present, and more likely sucrose~\cite{ueda_1976}) which may play a role of repellents preventing plasmodium from spreading inside the data set. Combination of these two properties causes \emph{P. polycephalum} to develop a network of protoplasmic tubes with major tube approximating convex hull. Exact mechanism of short-distance repelling is unclear and will be a topic of further investigations.   

Plasmodium of \emph{P. polycephalum} approximates shape of $\mathbf P$ by propagating simultaneously clock- and anti-clockwise (two branches). These branches fuse when meet one another. Thus the slime mould approximates shape of $\mathbf P$ in time $\Pi/2$, where $\Pi$ is a perimeter of the shape. The time can be reduced  further by $n$ by inoculating plasmodium in $n$ loci of space at the same. Indeed the slime mould based experimental computation of planar shape can not and will not compete with existing algorithms, however will contribute towards design of future parallel embedded processors made of non-linear chemical media, and also in control and navigation of amoeboid robots.


\begin{thebibliography}{}

\bibitem{adamatzky_1994} 
Adamatzky~A. Identification of Cellular Automata (Taylor \& Francis, 1994). Chapter 4, Sect. 4.7.2.

\bibitem{adamatzky_physarummachines}
Adamatzky~A. Physarum Machines: Making Computers from Slime Mould (World Scientific, 2010).

\bibitem{adamatzky_naturewissenschaften_2007}
Adamatzky A. 
Physarum machines: encapsulating reaction-diffusion to compute spanning tree. 
Naturwisseschaften 94 (2007) 975--980.


\bibitem{adamatzky_ppl_2007}
Adamatzky A.
Physarum machine: implementation of a Kolmogorov-Uspensky machine on a biological substrate.
Parallel Process Lett 17 (2007) 455--467.

\bibitem{adamatzky_gates}
Adamatzky~A.
Slime mould logical gates: exploring ballistic approach (2010).
\url{http://arxiv.org/abs/1005.2301}

\bibitem{adamatzky:valerian}
Adamatzky~A. Are slime moulds in love with insects? (2011), submitted.

\bibitem{akl_toussaint}
Akl~S.~G. and Toussaint~G.~T. 
A fast convex hull algorithm. 
Inform Process Lett 7 (1978) 219--222. 

\bibitem{chen_nakano_wada_2000}
Chen W., Nakano K. and Wada K.
Parallel algorithms for convex hull problems and their paradigm
IEICE Trans. Information and Syst. E83-D (2000) 519--529. 

\bibitem{clarridge_salomaa_2008}
Clarridge~A.~G. and Salomaa~K. 
An improved cellular automata based algorithm for the 45-convex hull problem.
J of Cellular Automata 5 (2010) 107--120.

\bibitem{day_tracey_1998}
Day A. M. and Tracey D.
Parallel implementations for determining the 2D convex hull
{\em Concurrency Practice and Experience} {\bf 10} (1998) 449--466.

\bibitem{deberg_2008}
de Berg~M., Cheong~O., van Kreveld~M., Overmars~M. 
Computational Geometry: Algorithms and Applications.
(Springer, 2008).

\bibitem{dehne:hassenklover}
Dehne F., Hassenklover A.L., Sack J.R. and Santoro N., Computation
geometry algorithms for the systolic screen {\em Algorithmica\/}
{\bf 6} (1991) 734--761.

\bibitem{edelsbrunner_1983}
Edelsbrunner~H., D. Kirkpatrick, R. Seidel, On the shape of a set of
points in the plane, IEEE Trans Inform Theor 29 (4)
(1983) 551–-559.


\bibitem{edelsbunner_1994}
Edelsbrunner~H., M\''{u}ke~E.~P.
Three-dimensional alpha shapes.
ACM Trans on Graphics 13 (1994) 43--72.

\bibitem{evans_mai_1985}
Evans D.J. and Mai S.-W.
Two parallel algorithms for the convex hull problem in a two dimensional space
{\em Parallel Comput} {\bf 2} (1985) 313--326.


\bibitem{field_noyes_1974}
Field~R.~J., Noyes~R.~M.
Oscillations in chemical systems. 
J. Chem. Phys. 60 (1974) 1877--1884.

\bibitem{jarvis_1973}
Jarvis~R. 
On the identification of the convex hull of a finite set of points in the plane. 
Inform Process Lett 2 (1973) 18--21.

\bibitem{hayashi:nakano}
Hayashi T., Nakano K. and Olariu S., Optimal parallel algorithms
for finding proximate points, with applications. 
IEEE Trans. Parall Dist Syst 9 (1998) 1153--1166.


\bibitem{graham_1972}
Graham~R.~L. 
An efficient algorithm for determining the convex hull of a finite planar set. 
Inform Process Lett 1 (1972) 132--133.


\bibitem{kamiya_1959}
Kamiya, N.
Protoplasmic streaming. Protoplasmatologia 8 (1959) 1–-199.
Cited by~\cite{tero_2007}.


\bibitem{kim:stojmenovic}
Kim C.E. and Stojmenovic I. Sequential and parallel approximate
convex hull algorithms  Computers and AI 14 (1995) 597--610.


\bibitem{nakagaki_yamada_1999}
Nakagaki~T., Yamada~H., Ueda~T.
Modulation of cellular rhythm and photoavoidance by
oscillatory irradiation in the Physarum plasmodium.
Biophys Chem 82 (1999) 23--28.

\bibitem{nakagaki_2000}
Nakagaki T., Yamada H., Ueda T. 
Interaction between cell shape and contraction pattern
in the {\it Physarum plasmodium}. Biophys Chemy 84 (2000) 195--204.


\bibitem{nakagaki_2001a}
Nakagaki T., Yamada H., and Toth A.,
 Path finding by tube morphogenesis in an amoeboid organism. 
Biophysical Chemistry 92 (2001) 47–-52.

\bibitem{orourke}
O'Rourke~J. Computational Geometry in C.
(Cambridge University, 2001).

\bibitem{preparata_hong}
Franco P. Preparata, S.J. Hong. Convex Hulls of Finite Sets of Points in Two and Three Dimensions, Commun. ACM
20 (1977) 87-–93.

\bibitem{preparata_shamos}
Preparata~F.~P. and Shamos~M.~I. 
Computational Geometry: An Introduction (Springer-Verlag New York Inc., 1985). 

\bibitem{schumann_adamatzky_2009}
Schumann~A. and Adamatzky~A.
Physarum spatial logic. In: Proc.
1th Int. Symp. on
Symbolic and Numeric Algorithms for Scientific Computing
(Timisoara, Romania, September 26-29, 2009).

\bibitem{shirakawa}
Shirakawa~T., Adamatzky~A., Gunji~Y.-P., Miyake~Y.
On simultaneous construction of Voronoi diagram and Delaunay triangulation by Physarum polycephalum.
Int. J. Bifurcation and Chaos 19 (2009 3109--3117.  

\bibitem{stephenson_2000}
Stephenson~S.~L. and Stempen~H.
Myxomycetes: A Handbook of Slime Molds.
(Timber Press, 2000).

\bibitem{tero_2007}
Tero~A., Kobayashi~R., Nakagaki~T.
A mathematical model for adaptive transport network in path finding
by true slime mold. J of Theor Biol 244 (2007) 553-–564.

\bibitem{torbey_akl_2008}
Torbey~S. and Akl~S.G., An exact and optimal local solution to the two-dimensional convex hull of arbitrary points problem, 
J of Cellular Automata (2008).

\bibitem{tsuda_2004}
Tsuda~S., Aono~M., Gunji~Y.-P.
Robust and emergent Physarum logical-computing. Biosystems 73 (2004) 45--55.

\bibitem{tyson_fife}
Tyson~J.~J., Fife~P.~C.
Target patterns in a realistic model of the Belousov–Zhabotinskii reaction.
J. Chem. Phys. 1980;73:2224--37.



\bibitem{ueda_1976}
Ueda T., Muratsugu M., Kurihara K. and Kobatake Y.
Chemotaxis in Physarum polycephalumnext term: Effects of chemicals on isometric tension of the plasmodial strand in relation to chemotactic movement. Experimental Cell Research 100 (1976) 337--344.

\end{thebibliography}
\end{document}